\documentclass[aps,prd,twocolumn,showkeys,nofootinbib]{revtex4-1}
\usepackage{epsfig,epsf,graphicx,graphics}
\usepackage{amsmath}
\usepackage{color}

\usepackage{lineno}
\usepackage [latin1]{inputenc}
\def\etal{{\it et al.}}

\begin{document}

\title{An Iterative Weighting Method to Apply ISR Correction to $e^+e^-$ Hadronic Cross-section Measurements}
\author{Wenyu Sun$^{a,b}$, Tong Liu$^{b}$, Maoqiang Jing$^{b}$, Liangliang Wang$^{b}$~\footnote{Email:llwang@ihep.ac.cn}, Bin Zhong$^{c}$~\footnote{Email:zhongb@njnu.edu.cn} and Weimin Song$^{a}$~\footnote{Email:weiminsong@jlu.edu.cn }  }
\affiliation{(a) College of Physics, Jilin University, Changchun 130012, China\\
             (b) Institute of High Energy Physics, Chinese Academy of Sciences, Beijing 100049, China\\
             (c) Department of Physics, Nanjing Normal University, Nanjing, Jiangsu 210023, China
             }

\begin{abstract}
   Initial state radiation (ISR) plays an important role in $e^+$$e^-$ collision experiments such as the BESIII. To correct the ISR effects in measurements of hadronic cross-sections of $e^+e^-$ annihilation, an iterative method that weights simulated ISR events is proposed here to assess the efficiency of event selection and the ISR correction factor for the observed cross-section. The simulated ISR events were generated only once, and the obtained cross-sectional line shape was used iteratively to weigh the same simulated ISR events to evaluate the efficiency and corrections until the results converge. Compared with the method of generating ISR events iteratively, the proposed weighting method provides consistent results, and reduces the computational time and disk space required by a factor of five or more, thus speeding-up $e^+e^-$ hadronic cross-section measurements.
\end{abstract}

\maketitle

\section{Introduction}
In $e^+$$e^-$ collision experiments, measurements of the hadronic cross-sections of $e^+e^-$ annihilation are important to search for new resonances and identify the reaction mechanisms. In addition to Born-level production, the initial state radiation (ISR) and the vacuum polarization (VP) process contribute to the observed hadronic cross-section that is measured directly. The ISR is a universal process in which one or more photons are emitted by an electron or a positron before they annihilate in $e^+$$e^-$ collision. The emitted photon or photons take part of energy from the electron or positron and reduce the center-of-mass (c.m.) energy of annihilation. Thus, the observed hadronic cross-section of $e^+e^-$ collision involves the line shape of the Born cross-section from the production threshold up to the nominal collision c.m.\ energy via the ISR. This effect needs to be considered in $e^+e^-$ hadronic cross-sectional measurements to compare the results with those of theoretical calculations.

Research on ISR correction has a long history. The calculation of the contribution of the ISR is an application of the Feynman rules for quantum electrodynamics (QED) \cite{Blumlein:2020jrf,Blumlein:2019pqb,Blumlein:2019srk}. A large number of studies have described ISR corrections in theory in different ways and precisions \cite{ISR:intro0,ISR:intro1,ISR:intro2,ISR:intro3,ISR:intro4,ISR:intro5}. This has led to the development of dedicated, high-precision Monte Carlo (MC) generators, such as the MCGPJ, PHOKHARA, TOPAZO, and ZIFITTER for $e^+$$e^-$ collision experiments in recent years \cite{ISR:intro00,Montagna:1998kp,Arbuzov:2005ma}.

\begin{figure}[!htbp]
\begin{center}
\includegraphics[width=0.38\textwidth]{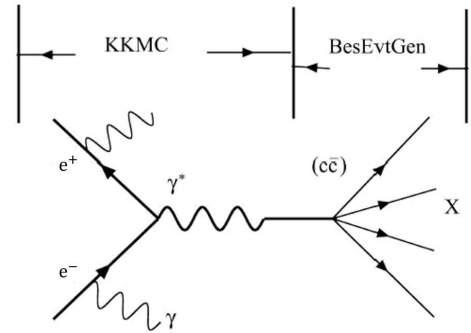}
\end{center}
    \vskip -0.7cm
    \caption{Illustration of the BESIII generator framework.}\label{Fig:genframe}
\end{figure}

The BESIII is one such vehicle for $e^+$$e^-$ collision experiments, and is used as an example in the discussion here. The BESIII detector is a magnetic spectrometer at the Beijing Electron Positron Collider (BEPCII) \cite{BESIII:intro}. The advanced design of BESIII allows it to take advantage of the high luminosity delivered by BEPCII and collect large data samples at the $\tau$-charm energy region, which is between the perturbative and non-perturbative regimes of quantum chromodynamics (QCD) from 2 ~GeV to 4.7~GeV.
At BESIII, KKMC + BesEvtGen is the most commonly used generator framework, and is used to generate MC events for a process of $e^+$ $e^-$ annihilation into final states $X$, as shown in Figure~\ref{Fig:genframe}. The KKMC is used to generate the intermediate states by considering the ISR and the beam energy spread. The BesEvtGen was developed based on the EvtGen generator, and is used to generate the final states of the intermediate state decays with a final state radiation (FSR) \cite{bes3:KKMC}. The responses of the BESIII detector to the generated events are simulated by a Geant4-based algorithm consisting of a description of the detector, digitization, and backgrounds mixing ~\cite{Deng:2007zzb}. The simulated ISR events are reconstructed and selected as data to estimate efficiency and the corrections that are needed for the measurement.

This paper proposes an iterative method of weighting the simulated ISR events to apply ISR correction. The procedure is explained in detail, and three line shapes are used as examples to validate it.

\section{ISR CORRECTION PROCEDURE}
The observed cross-section $\sigma^\mathrm{obs}$ of a process $e^+e^-\to \mathrm{hadrons}$ can be obtained experimentally by
\begin{equation}
\sigma^\mathrm{obs}=\frac{N^\mathrm{obs}}{\epsilon\cdot\mathcal{L}}, \label{eq:cobs}
\end{equation}
where $N^\mathrm{obs}$ is the number of observed signals, $\epsilon$ is efficiency, and $\mathcal{L}$ is the integrated luminosity of the data sample.

As mentioned above, the observed cross-section includes effects from the ISR and VP. We focus on ISR correction in this article. The observed cross-section can be corrected to a cross-section without ISR contribution, called the ``dressed cross-section" by
\begin{eqnarray}
\sigma^{\mathrm{dressed}}=\frac{\sigma^\mathrm{obs}}{(1+\delta)}=\frac{N^\mathrm{obs}}{\epsilon\cdot\mathcal{L}\cdot(1+\delta)}, \label{eq:cisr}
\end{eqnarray}
where $1+\delta$ is the ISR correction factor. Then the ``dressed cross-section" can be corrected to the ``Born cross-section" by
\begin{eqnarray}
\sigma^{\mathrm{Born}}=\frac{\sigma^\mathrm{dressed}}{f_\mathrm{vp}}, \label{eq:csBorn}
\end{eqnarray}
where $f_{vp}$ is the VP correction factor.

The efficiency $\epsilon$ and the ISR correction factor $1+\delta$ can be evaluated by simulating ISR events usually generated by the KKMC at BESIII. To generate these events, a line shape of the dressed cross-section as a function of energy is an essential input. For some $e^+e^-\to$ hadronic cross-section measurements, nothing or very little is known about the line shape of the cross-section before the measurement is performed. A line shape of the cross-section $\sigma^\mathrm{dressed}_0(E_\mathrm{cm})$ (such as the Breit--Wigner function or a flat line) is typically assumed in the KKMC to generate MC events as a starting point to estimate the efficiency and the ISR correction factor at each energy point, where $E_\mathrm{cm}$ is the c.m.\ energy of $e^+e^-$ annihilation. The dressed cross-section can then be roughly obtained at each energy point by using Equation~\ref{eq:cisr}. Following this, the line shape of the estimated dressed cross-section is used to re-valuate efficiency and the ISR correction factor. The procedure needs to be repeated several times, normally more than five times, until all results converge. In other words, the final dressed cross-sections are measured in an iterative way.

\subsection{Iterative MC Generation Method}
\begin{figure}[!htbp]
\begin{center}
\includegraphics[width=0.49\textwidth]{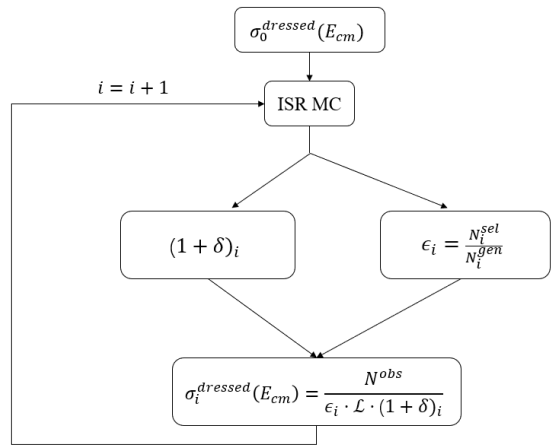}
\end{center}
\caption{Flowchart of the iterative MC generation method}\label{Fig:oldIterframe}
\end{figure}

To evaluate the efficiency $\epsilon$ and the ISR correction factor $1+\delta$ precisely, a natural way is to re-generate the MC events with ISR effect, with the KKMC taking the new line shape as input. The iterative procedure to regenerate MC events with ISR effect, called the ``iterative MC generation method" is presented in Figure \ref{Fig:oldIterframe}.

As mentioned above, a line shape $\sigma^\mathrm{dressed}_0(E_\mathrm{cm})$ of the cross-section is assumed to generate signal MC samples with ISR effect by the KKMC to start the iteration. In the $i$-th iteration ($i=1, 2, 3, ...$), the ISR correction factor $(1+\delta)_i$ is obtained directly from the KKMC; after the reconstruction and event selection of the MC sample, efficiency is obtained by $\epsilon_i={N^\mathrm{sel}_i}/{N^\mathrm{gen}_i}$, where $N^\mathrm{sel}_i$ is the number of events after selection and $N^\mathrm{gen}_i$ is the generated number of events. Then, the dressed cross-section after $i$-th iteration is $\sigma^\mathrm{dressed}_i=\frac{N^\mathrm{obs}}{\epsilon_i\cdot\mathcal{L}\cdot(1+\delta)_i}$. This operation is performed for all energy points to obtain the line shape of the dressed cross-section after the $i$-th iteration $\sigma^\mathrm{dressed}_i(E_\mathrm{cm})$, where this can be used as input for the KKMC to generate signal MC samples with ISR effect in the next (i.e., $(i+1)$-th) iteration. The iterations can stop if the results converge (for example, if the relative difference in $\sigma^\mathrm{dressed}(E_\mathrm{cm})$ between a given iteration and previous one is smaller than a certain quantity).

This method is widely used in many cross-sectional measurements at the BESIII, such as $e^{+}e^{-}\rightarrow\pi^{+}\pi^{-}h_{c}$, $e^{+}e^{-}\rightarrow\pi^{+}\pi^{-}J/\psi$, and $e^{+}e^{-}\rightarrow\pi^{+}\pi^{-}\psi(3686)$ ~\cite{bes3:p1,bes3:p2,bes3:p3,bes3:p4,bes3:p5,bes3:p6,bes3:p7}.
The iterative MC generation method works well but requires a long computational time and a large amount of disk space to execute multiple rounds of MC generation, the Geant4 simulation of the BESIII detector, reconstruction, and event selection, especially when a large number of energy points are used. The measurements are slowed further if the related systematic uncertainty is considered.

\subsection{Iterative MC Weighting Method}
\begin{figure}[!htbp]
\begin{center}
\includegraphics[width=0.49\textwidth]{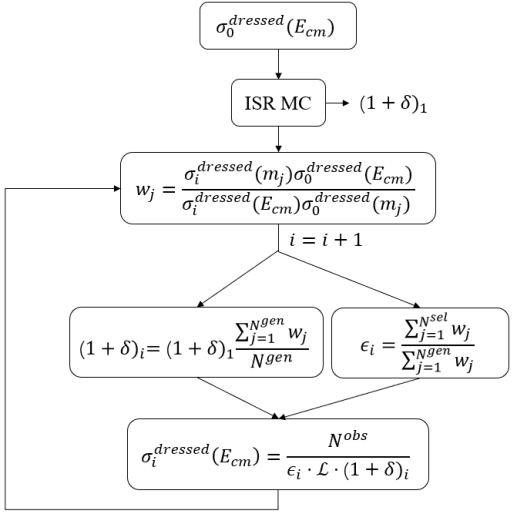}
\end{center}
\caption{Flowchart of the iterative MC weighting method}\label{Fig:newIterframe}
\end{figure}

A more efficient iterative method is proposed here, in which the MC events with ISR effect are generated, reconstructed, and selected only once at each energy point. The starting point is similar: from these MC events with ISR effect generated by the KKMC using the assumed line shape $\sigma^\mathrm{dressed}_0(E_\mathrm{cm})$ of the cross-section, the rough ISR correction factor $(1+\delta)_1$ and efficiency $\epsilon_1=N^\mathrm{sel}/N^\mathrm{gen}$ are obtained, where $N^\mathrm{sel}$ is the number of events after selection and $N^\mathrm{gen}$ is the generated number of events. Then, the rough cross-section is obtained by $\sigma^\mathrm{dressed}_1=\frac{N^\mathrm{obs}}{\epsilon_1\cdot\mathcal{L}\cdot(1+\delta)_1}$. In subsequent iterations, the same MC events with ISR effect are weighted according to the line shape of the cross-section from the previous iteration to calculate the ISR correction factor, the efficiency of event selection \footnote{in some cases, $N^\mathrm{obs}$ also needs to be re-extracted if it depends on the ISR effects in MC}, and, subsequently, the cross-section at each energy point. The flowchart of this iterative MC weighting method is shown in Figure~\ref{Fig:newIterframe}.

For a specific MC event $j$ generated at  $E_\mathrm{cm}$, the corresponding differential cross-section is proportional to $F(m_{j}, E_\mathrm{cm})\sigma^\mathrm{dressed}_{0}(m_{j})$~\cite{Actis:2010gg}, where $F(m_j, E_\mathrm{cm})$ is the radiation function and $m_j$ is the invariant mass of the hadrons (FSR included) in this event. Similarly, if this event is generated according to the cross-sectional line shape $\sigma^\mathrm{dressed}_{i}(E_\mathrm{cm})$ instead, the corresponding differential cross-section should be proportional to $F(m_{j}, E_\mathrm{cm})\sigma^\mathrm{dressed}_{i}(m_{j})$. Thus, the ratio of the two differential cross-sections is
\begin{equation}\label{W_factor}
r_{j}=\frac{F(m_{j}, E_\mathrm{cm})\sigma^\mathrm{dressed}_{i}(m_{j})}{F(m_{j}, E_\mathrm{cm})\sigma^\mathrm{dressed}_{0}(m_{j})}=\frac{\sigma^\mathrm{dressed}_{i}(m_{j})}{\sigma^\mathrm{dressed}_{0}(m_{j})}.
\end{equation}

 For the MC sample generated at $E_\mathrm{cm}$ with the dressed cross-sectional line shape $\sigma^\mathrm{dressed}_{0}(E_\mathrm{cm})$, the corresponding luminosity is $\mathcal{L}_\mathrm{MC}=N^\mathrm{gen}/\sigma^\mathrm{obs}_{0}(E_\mathrm{cm})$, where $\sigma^\mathrm{obs}_{0}(E_\mathrm{cm})$ is the observed cross-section. If  a different cross-section line shape $\sigma^\mathrm{dressed}_{i}(E_\mathrm{cm})$ is considered, event $j$ should be generated $r_j$ times instead of once to make the total number of events corresponding to the same luminosity $\mathcal{L}_\mathrm{MC}$. So, the ratio of the total observed cross-sections with the two cross-sectional line shapes can be calculated using these MC events in terms of $r_{j}$:

\begin{equation}\label{TOT_ob_dif}
\frac{\sigma^\mathrm{obs}_{i}(E_\mathrm{cm})}{\sigma^\mathrm{obs}_{0}(E_\mathrm{cm})}=\frac{\mathcal{L}_\mathrm{MC}\cdot\sigma^\mathrm{obs}_{i}(E_\mathrm{cm})}{\mathcal{L}_\mathrm{MC}\cdot\sigma^\mathrm{obs}_{0}(E_\mathrm{cm})}=\frac{\sum^{N^{gen}}_{j=1}r_{j}}{\sum ^{N^\mathrm{gen}}_{j=1}1}=\frac{\sum^{N^\mathrm{gen}}_{j=1}r_{j}}{N^\mathrm{gen}}.
\end{equation}

Then, starting from the definition of the ISR correction factor, and using Equation~\ref{TOT_ob_dif}, the ISR correction factor $(1+\delta)_{i+1}$ for the cross-sectional line shape $\sigma^\mathrm{dressed}_{i}(E_\mathrm{cm})$ can be related to the original ISR correction factor $(1+\delta)_{1}$ via the same MC events with ISR effect as follows:
\begin{equation}\label{ini_rad_new}
\begin{split}
(1+\delta)_{i+1}&=\frac{\sigma^\mathrm{obs}_{i}(E_\mathrm{cm})}{\sigma^\mathrm{dressed}_{i}(E_\mathrm{cm})}\\
&=\frac{\sigma_{0}^\mathrm{obs}(E_\mathrm{cm})\sum^{N^\mathrm{gen}}_{j=1}r_{j}/N^\mathrm{gen}}{\sigma^\mathrm{dressed}_{i}(E_\mathrm{cm})}\\
&=\frac{\sigma_{0}^\mathrm{obs}(E_\mathrm{cm})\sum^{N^\mathrm{gen}}_{j=1}\frac{\sigma^\mathrm{dressed}_{i}(m_{j})}{\sigma^\mathrm{dressed}_{0}(m_{j})}}{\sigma^\mathrm{dressed}_{i}(E_\mathrm{cm})N^\mathrm{gen}}\\
&=\frac{\sigma_{0}^\mathrm{obs}(E_\mathrm{cm})}{\sigma^\mathrm{dressed}_{0}(E_\mathrm{cm})} \frac{\sum^{N^\mathrm{gen}}_{j=1}\frac{\sigma^\mathrm{dressed}_{i}(m_{j})\sigma^\mathrm{dressed}_{0}(E_\mathrm{cm})}{\sigma^\mathrm{dressed}_{i}(E_\mathrm{cm})\sigma^\mathrm{dressed}_{0}(m_{j})}} {N^\mathrm{gen}}\\
&=(1+\delta)_{1}\frac{\sum^{N^\mathrm{gen}}_{j=1}
\frac{\sigma^\mathrm{dressed}_{i}(m_{j})\sigma^\mathrm{dressed}_{0}(E_\mathrm{cm})}{\sigma^\mathrm{dressed}_{i}(E_\mathrm{cm})\sigma^\mathrm{dressed}_{0}(m_{j})}
}{N^\mathrm{gen}}.
\end{split}
\end{equation}
Thus, if a weight
\begin{equation}\label{equ_weight}
w_j\equiv\frac{\sigma^\mathrm{dressed}_{i}(m_{j})\sigma^\mathrm{dressed}_{0}(E_\mathrm{cm})}{\sigma^\mathrm{dressed}_{i}(E_\mathrm{cm})\sigma^\mathrm{dressed}_{0}(m_{j})}
\end{equation}
is calculated for MC event $j$, the ratio of the sum of the weights over $N^\mathrm{gen}$ can be used to convert $(1+\delta)_{1}$ into $(1+\delta)_{i+1}$. If the original cross-sectional line shape used in ISR MC generation is flat, i.e., $\sigma^\mathrm{dressed}_{0}(E_\mathrm{cm})$ is a constant, the weight can be simplified as $w_j=\frac{\sigma^\mathrm{dressed}_{i}(m_{j})}{\sigma^\mathrm{dressed}_{i}(E_\mathrm{cm})}$.

The efficiency of event selection $\epsilon_{i+1}$ for the dressed cross-sectional line shape $\sigma^\mathrm{dressed}_i(E_\mathrm{cm})$ can also be estimated with the same MC events with ISR effect in terms of weight by
\begin{equation}\label{effiST}
  \epsilon_{i+1}=\frac{\sum^{N^\mathrm{sel}}_{j=1}w_{j}}{\sum_{j=1}^{N_\mathrm{gen}}w_{j}}.
\end{equation}

Except for the production of MC events with ISR effect at the beginning, in subsequent iterations, the MC weighting method involves only a loop of the MC events with ISR effect to calculate the weights. These weights are then summed to calculate the ISR correction factor, efficiency, and, subsequently, the dressed cross-sections. This method therefore requires a limited amount of computing time and disk space during the iterations to significantly speed-up the entire procedure.

\section{Validation of the MC Weighting Method}
To validate it, we compared the obtained ISR correction factor and efficiency with those obtained using the MC generating method.

\begin{figure*}[!htbp]
\begin{center}
\includegraphics[width=0.32\textwidth]{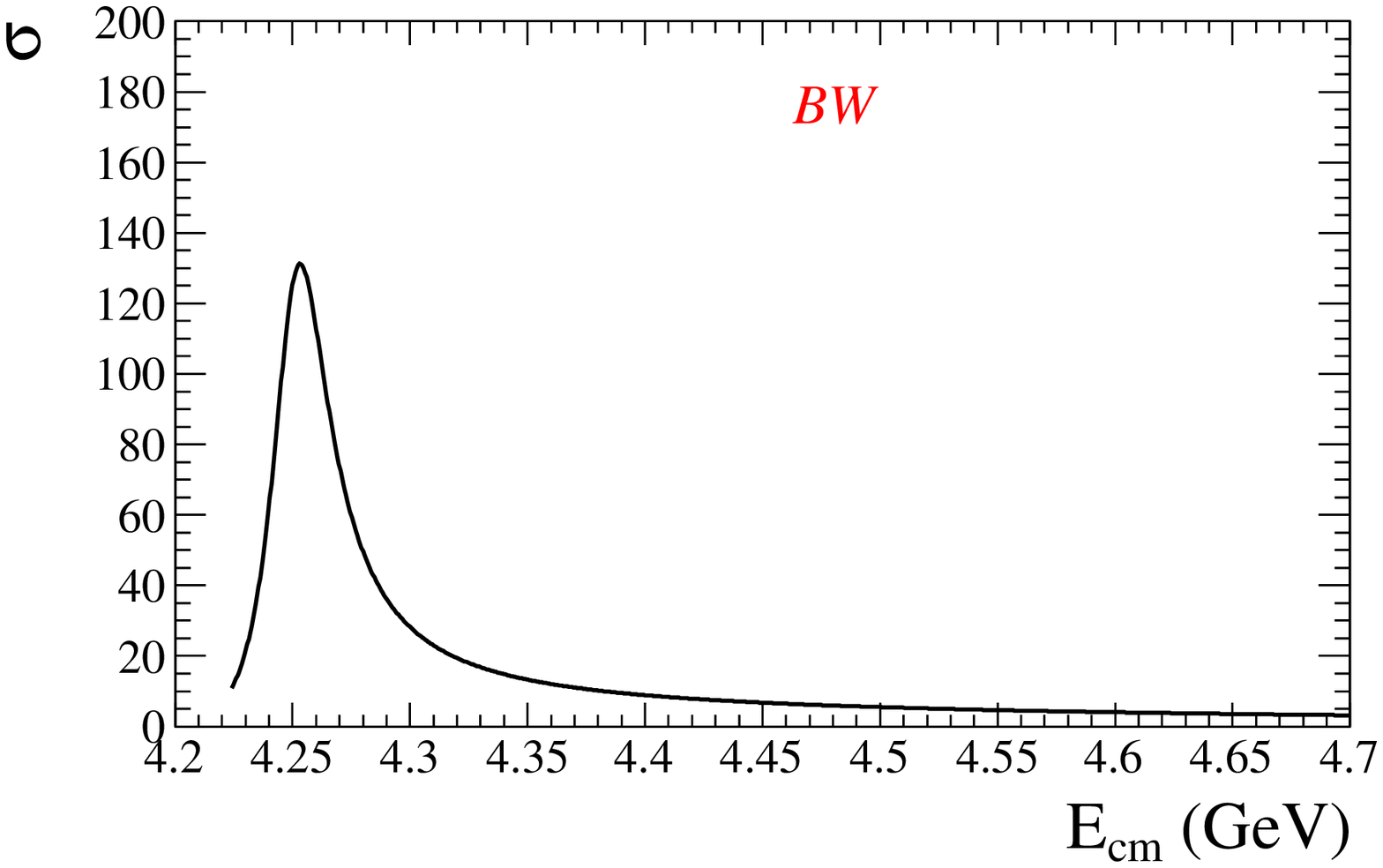}
\includegraphics[width=0.32\textwidth]{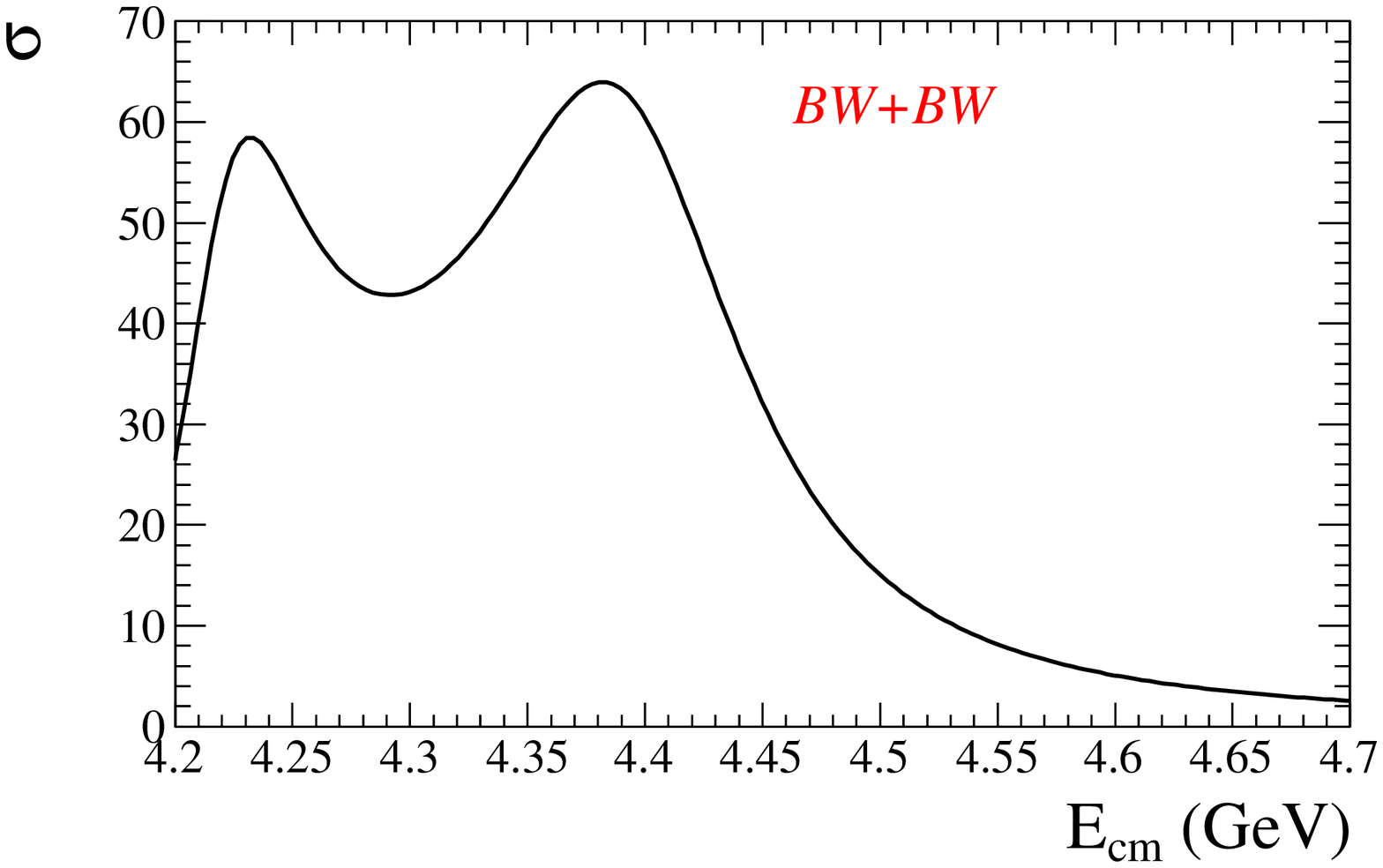}
\includegraphics[width=0.32\textwidth]{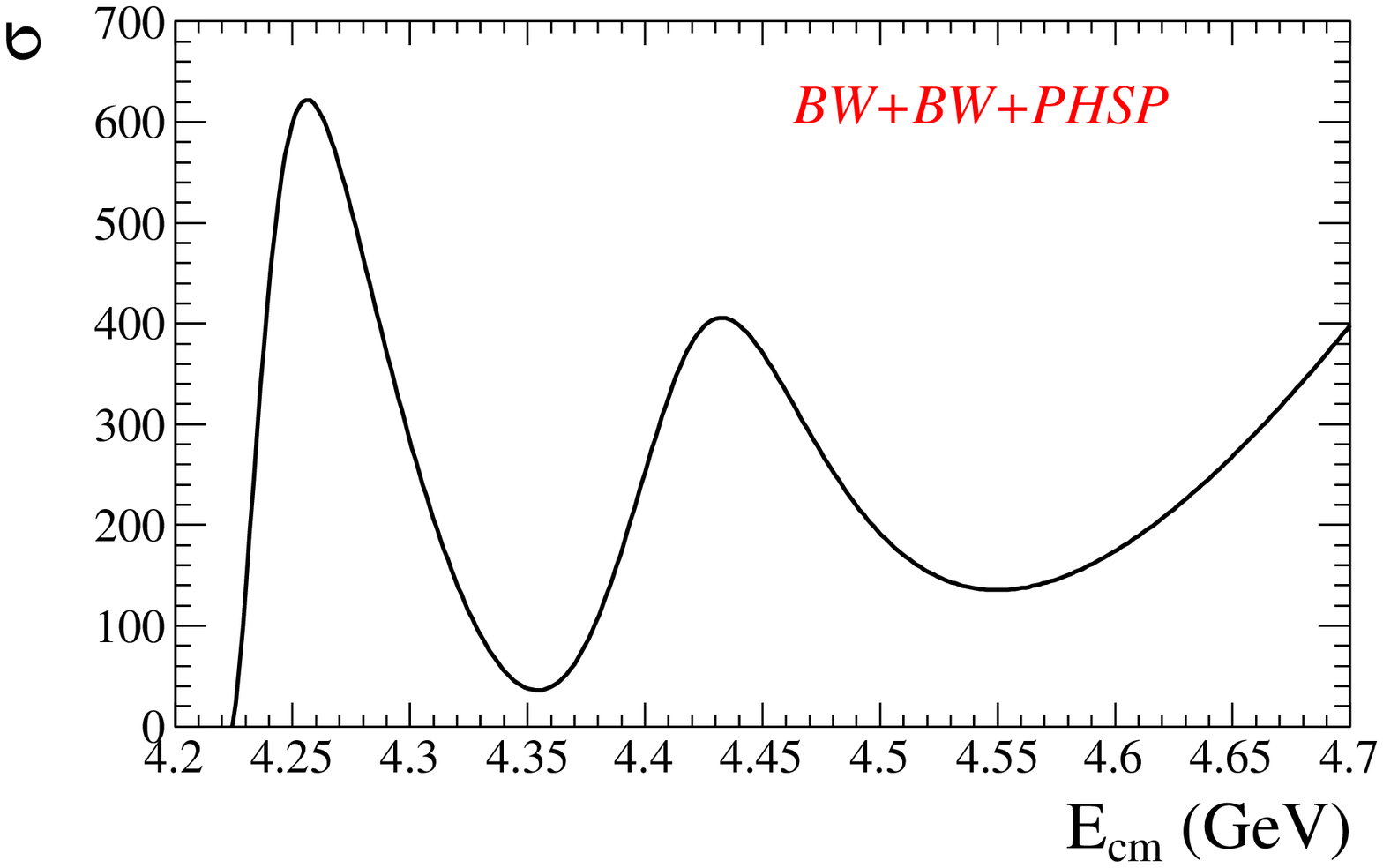}
\caption{Three line shapes to validate the MC weighting method. From left to right, one Breit--Wigner function, two Breit--Wigner functions, and two Breit--Wigner functions plus a phase space function.}\label{Fig:lineshape}
\end{center}
\end{figure*}

Three line shapes (arbitrarily selected according to the cross-sectional fitting habit) were used as examples to test the method: 1) one--Breit--Wigner function 2) two--Breit--Wigner functions, and 3) two--Breit--Wigner functions plus a phase space function. The line shapes are shown in Figure~\ref{Fig:lineshape}.

MC samples with ISR effect using a flat line shape were generated (including the simulation, reconstruction, and event selection). Following the MC weighting method, the three line shapes were used to calculate the corresponding weights, ISR correction factors, and efficiencies of event selection. They were also used to directly generate the corresponding MC samples with ISR effect (i.e., with the so-called ``MC generation method") to obtain the ISR correction factors and efficiencies of event selection.

\begin{figure*}[!htbp]
\begin{center}
\includegraphics[width=0.32\textwidth]{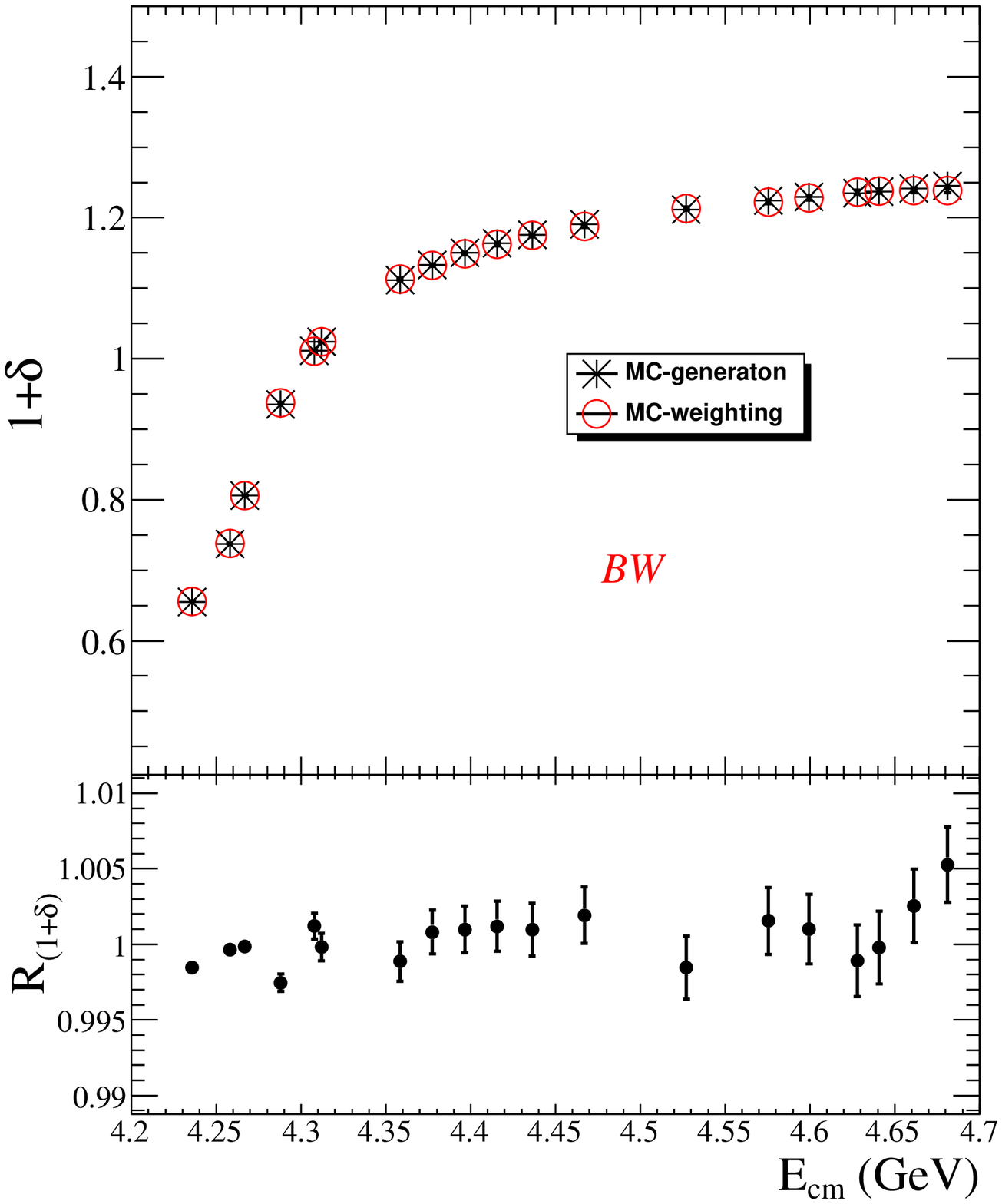}
\includegraphics[width=0.32\textwidth]{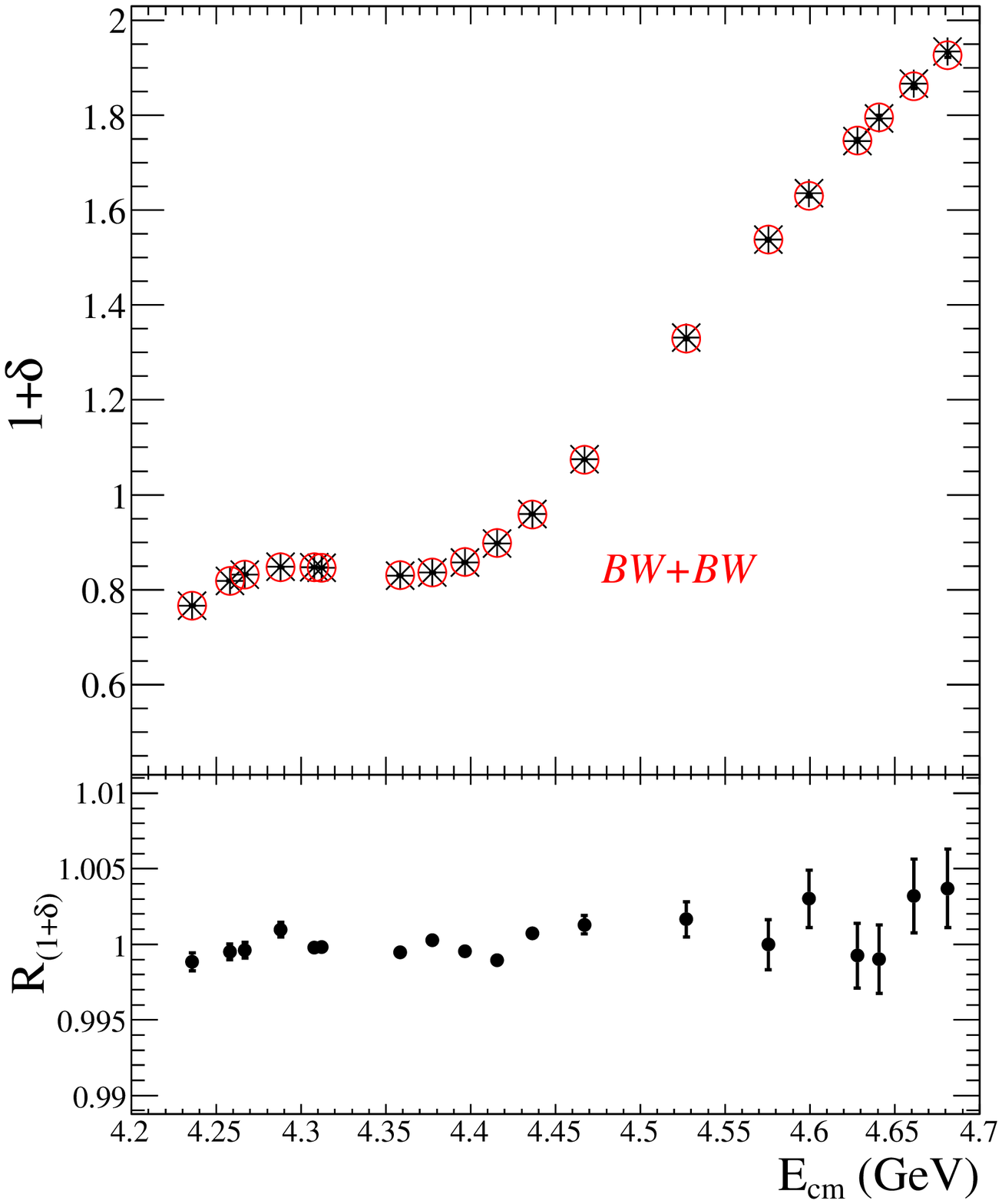}
\includegraphics[width=0.32\textwidth]{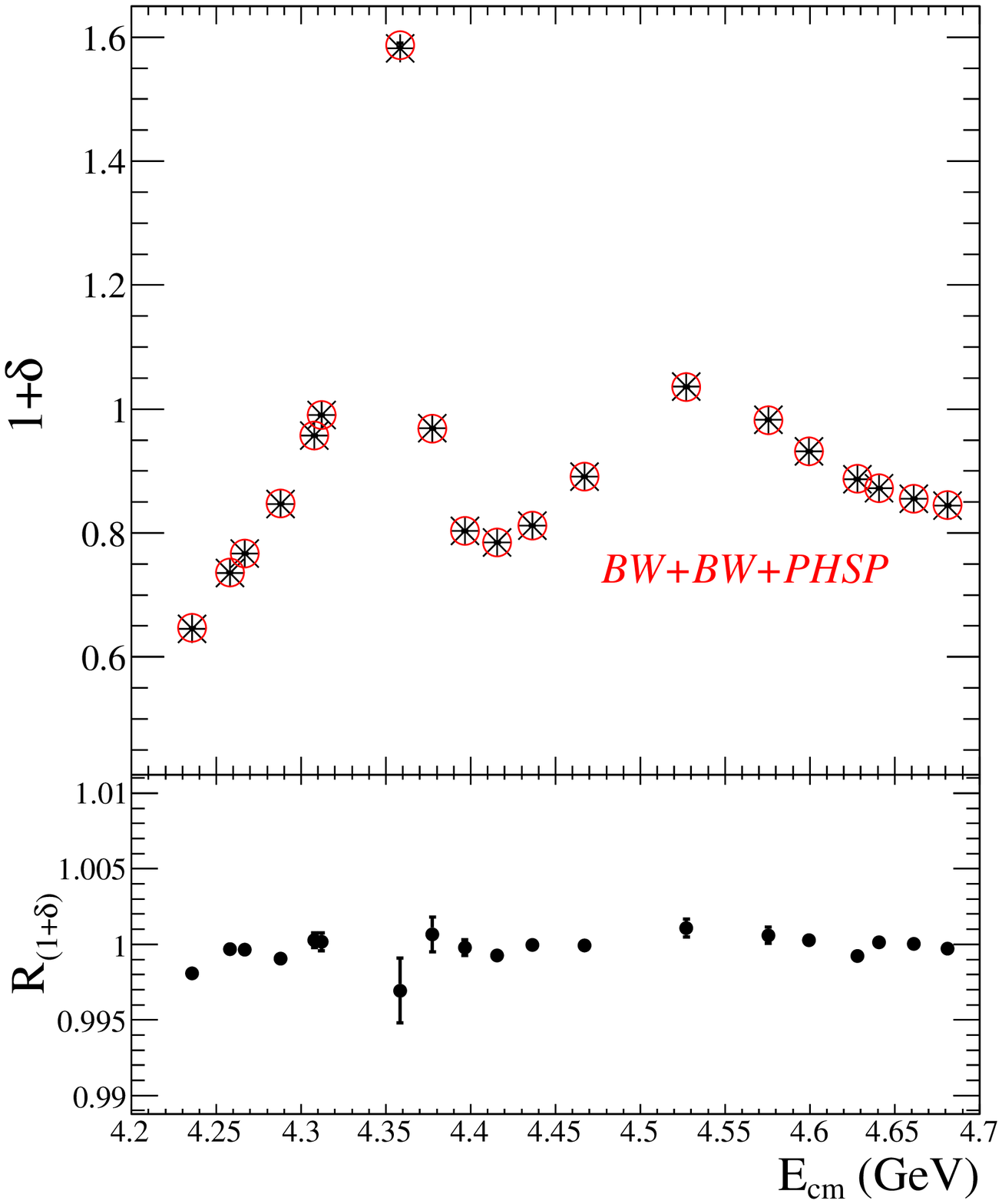}
\includegraphics[width=0.32\textwidth]{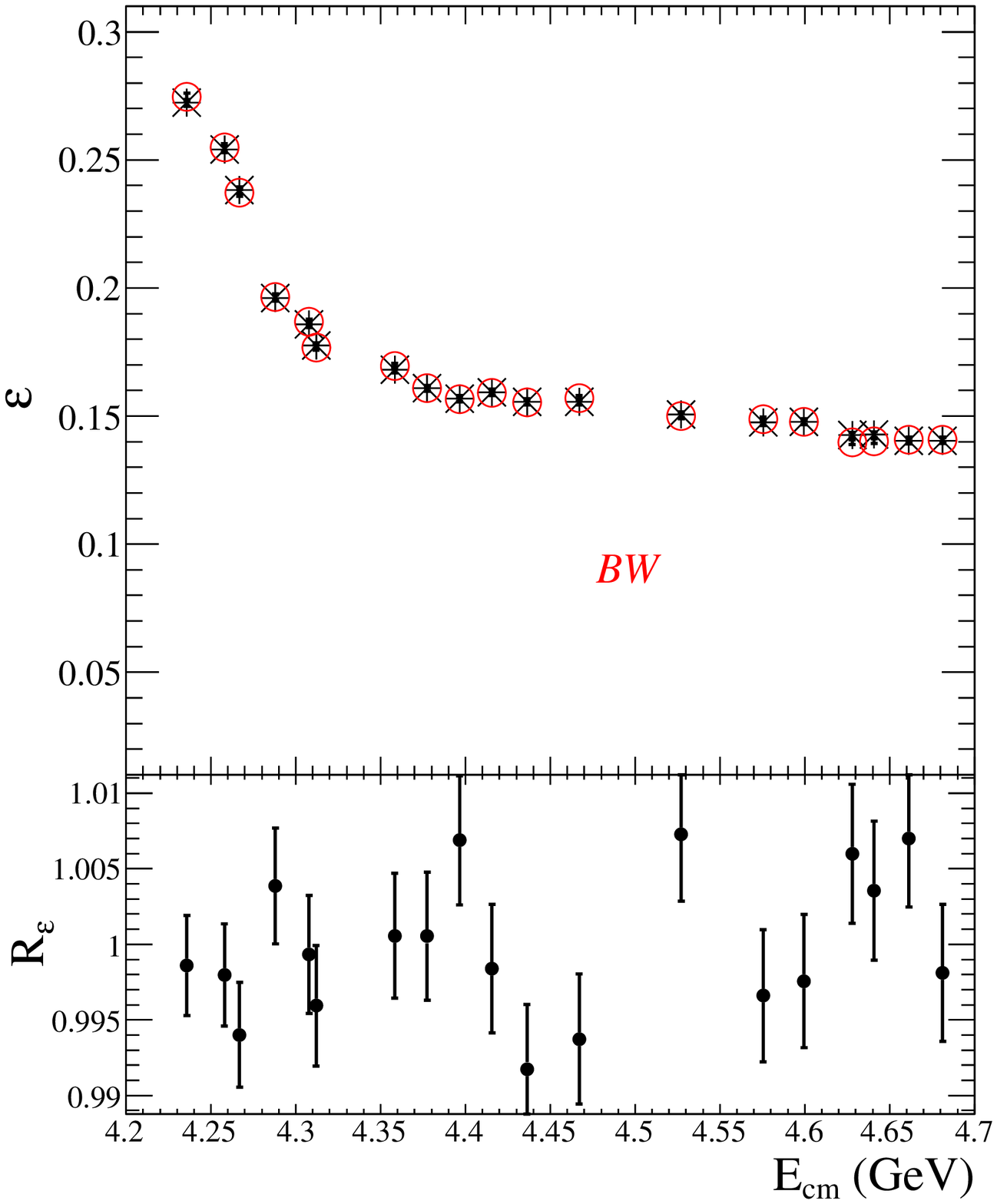}
\includegraphics[width=0.32\textwidth]{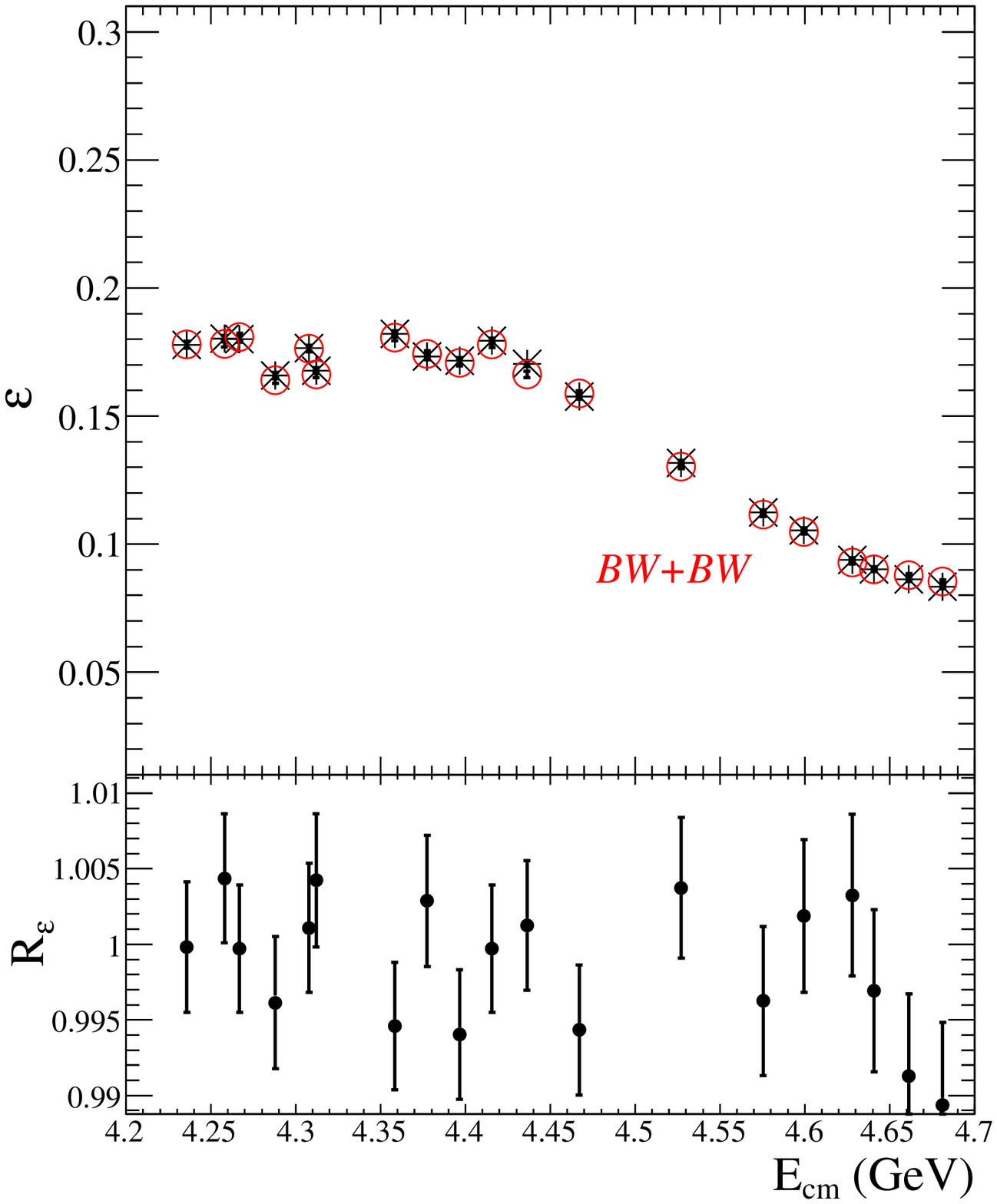}
\includegraphics[width=0.32\textwidth]{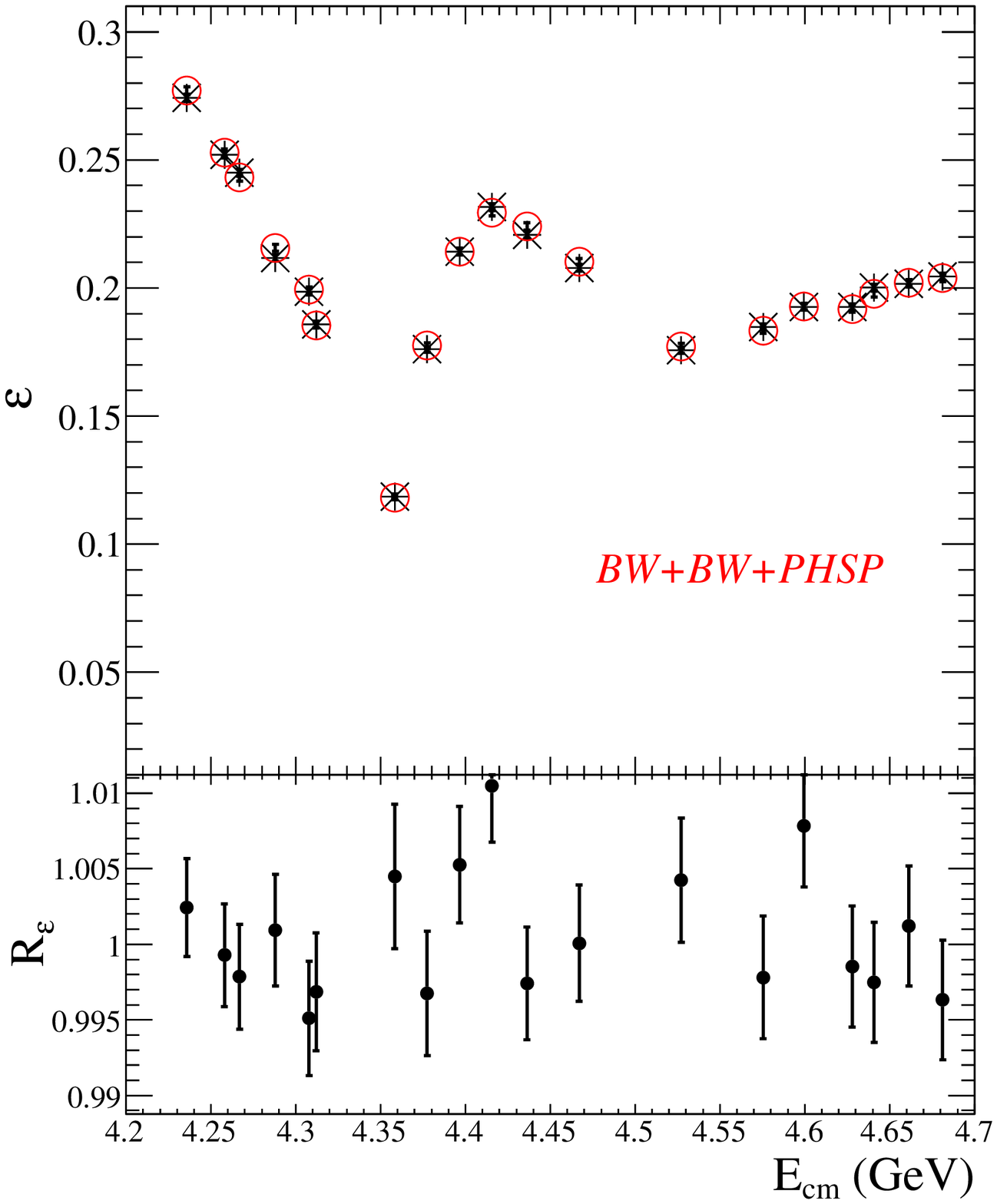}
\includegraphics[width=0.32\textwidth]{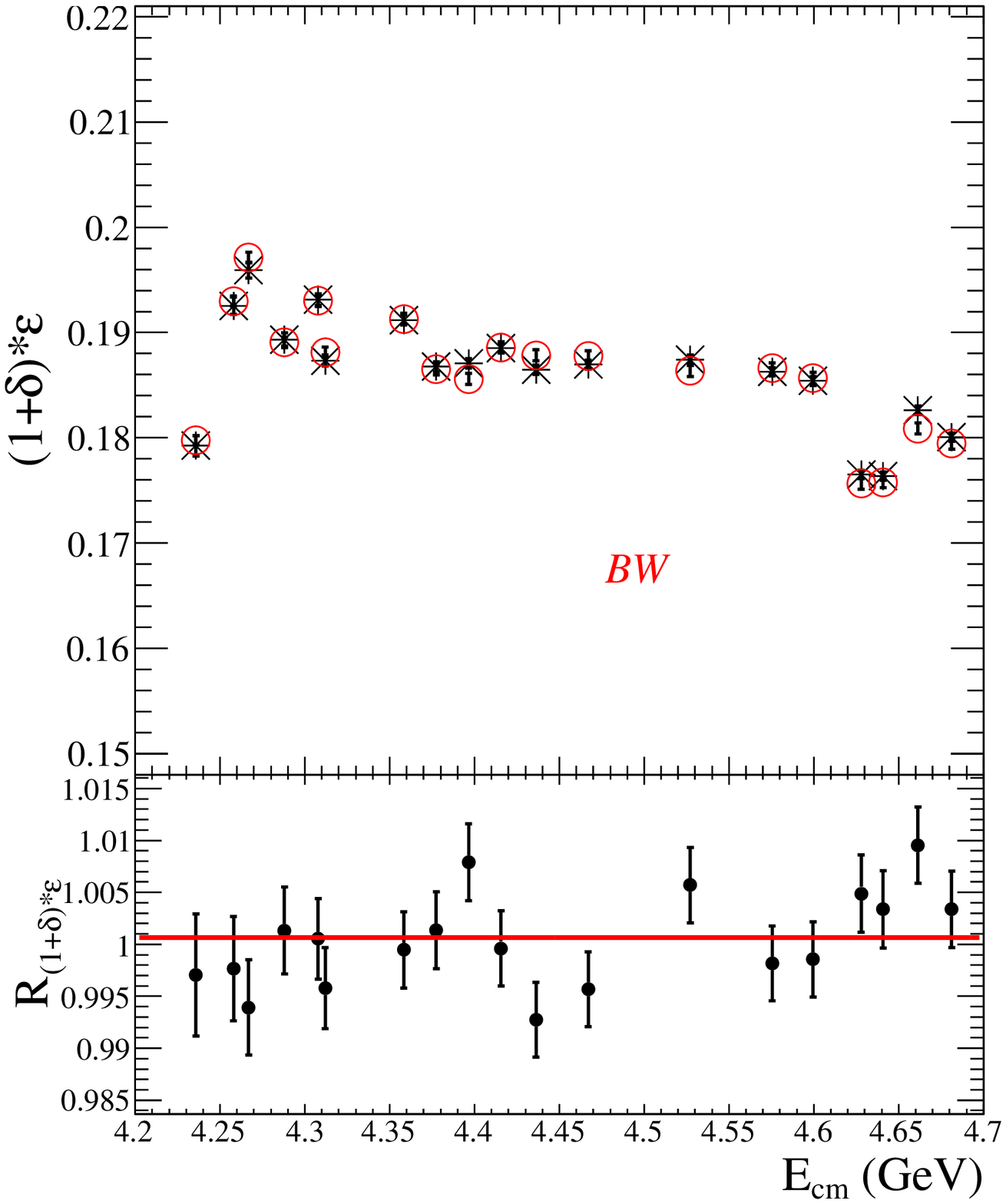}
\includegraphics[width=0.32\textwidth]{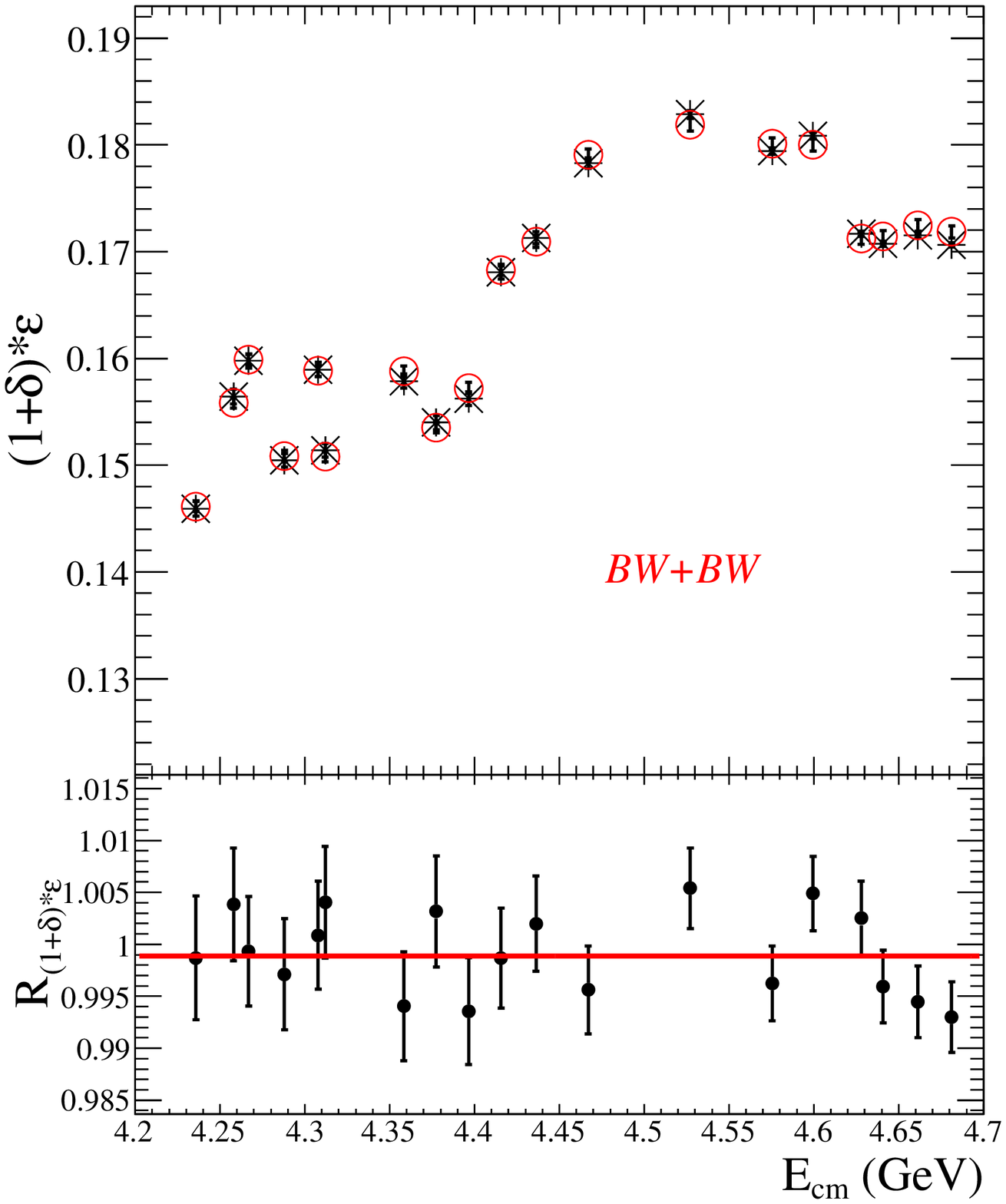}
\includegraphics[width=0.32\textwidth]{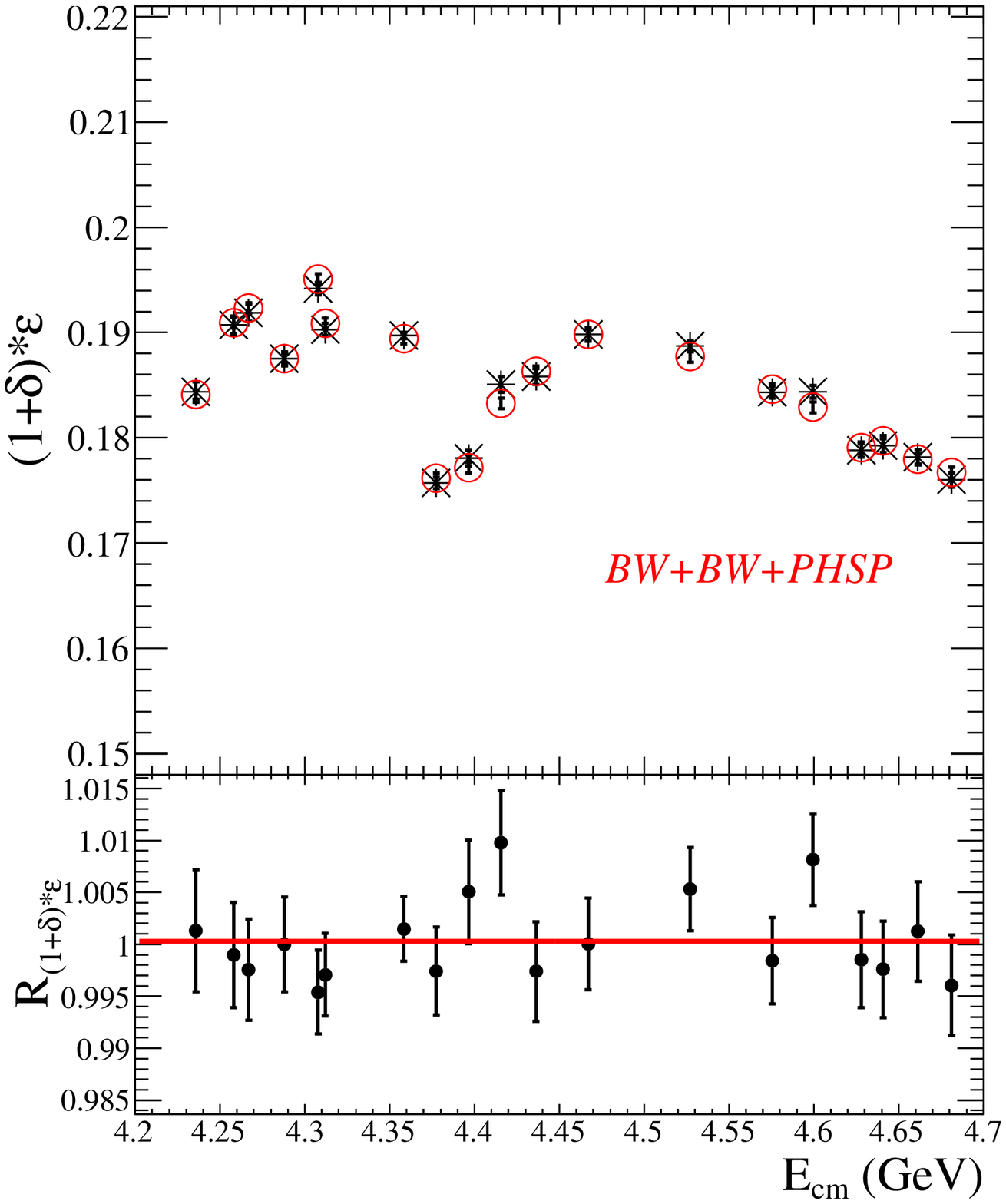}
\caption{A comparison of ISR correction factors (top), efficiencies of event selection (middle), and products of the two (bottom) as functions of energy for line shapes of the one Breit--Wigner function (the first column), two Breit--Wigner functions (the second column), and two Breit--Wigner functions plus a phase space (the third column) between the MC weighting method (red) and the MC generation method (black). The ratios between the two methods are shown in the subplots.  Fits to a constant are performed for the ratios of $(1+\delta)\times \epsilon$, and the fitting results are shown by red lines, which are $1.0006\pm0.0009$ (left), $0.9989\pm0.0010$ (middle), $1.0003\pm0.0010$ (right). The values of $\chi^{2}$/$\mathrm{NDF}$ from the fits are 26/18 (left), 18/18 (middle) and 14/18 (right), respectively. }\label{Fig:BW_compare}
\end{center}
\end{figure*}

In the tests performed here, 500000  MC events with ISR effect were generated at each energy point with each line shape. The ISR correction factors, efficiencies of event selection,  the products of ISR correction factors and efficiencies of event selection using the two methods as well as their ratios are shown in Figure~\ref{Fig:BW_compare}. The fluctuations of ratios around 1 are compatible with the statical errors.
The ratios of the products of ISR correction factors and efficiencies of event selection ($R_{(1+\delta)*\epsilon}$) are fitted to constants separately to get the averaged values which are consistent with $1$ within an error of $\sim0.1\%$.
So the two methods can produce the consistent results and no systematic uncertainty is necessary to be assigned to the MC weighting method. The uncertainty at each energy point is due to the statistics of the MC sample being used.


\vspace{0.8cm}

\section{Conclusion}

An iterative MC weighting method was proposed here to evaluate ISR correction factors and efficiencies of event selection in $e^+e^-$ hadronic cross-section measurements in $e^+e^-$ collision experiments, like those performed at the BESIII. The MC weighting method is identical to the MC generation method as proved numerically by three different line-shape examples. While as the MC samples with ISR effect are produced only once in this new procedure, the iterative MC weighting method can reduce the required computation time and disk space by a factor of more than five (depending on the number of iterations) compared with the iterative MC generation method.

\section*{Acknowledgments}
The material presented in this paper is by the authors alone, and has not been reviewed by the BESIII collaboration. However, we thank our colleagues for allowing us to make use of the BESIII software environment. This work was supported by the National Natural Science Foundation of China, under Grant Nos.~U1732105, U1632106, 11947415 and 12047569 and by the Research Foundation for Advanced Talents of Nanjing Normal University, under Grant No.~2014102XGQ0085. We also appreciate valuable discussions with Dr. Kai Zhu, Dr. Yuping Guo, and Prof. Changzheng Yuan.



\begin{references}
\frenchspacing


\bibitem{Blumlein:2020jrf} J.~Bl\"umlein, A.~De Freitas, C.~Raab and K.~Sch\"onwald, Nucl. Phys. B, 956: 115055 (2020)

\bibitem{Blumlein:2019pqb} J.~Bl\"umlein, A.~De Freitas, C.~G.~Raab and K.~Sch\"onwald, Phys. Lett. B, 801: 135196 (2020)

\bibitem{Blumlein:2019srk} J.~Bl\"umlein, A.~De Freitas, C.~G.~Raab and K.~Sch\"onwald, Phys. Lett. B, 791: 206 (2019)



\bibitem{ISR:intro0} G. Bonneau and F. Martin, Nucl. Phys. B, 27: 381 (1971)

\bibitem{ISR:intro1} D. R. Yennie, Phys. Rev. Lett., 34: 239 (1975)

\bibitem{ISR:intro2} V. N. Baier \etal, Phys. Rep., 78: 293 (1981)

\bibitem{ISR:intro3} F. A. Berends and R. Kleiss, Nucl. Phys. B, 178: 141 (1981)

\bibitem{ISR:intro4} E. A. Kuraev and V. S. Fadin, Sov. J. Nucl. Phys., 41: 466 (1985)

\bibitem{ISR:intro5} O. Nicrosini and L. Trentadue, Phys. Lett. B, 196: 551 (1987)

\bibitem{ISR:intro00} Ablinger, J. and Bl\"umlein, J. and De Freitas, A. and Sch\"onwald, K., Nucl. Phys. B, 955: 115045 (2020)


\bibitem{Montagna:1998kp} G.~Montagna, O.~Nicrosini, F.~Piccinini and G.~Passarino, Comput. Phys. Commun. 117: 278 (1999)

\bibitem{Arbuzov:2005ma} A.~B.~Arbuzov, M.~Awramik, M.~Czakon, A.~Freitas, M.~W.~Grunewald, K.~Monig, S.~Riemann and T.~Riemann, Comput. Phys. Commun, 174: 728 (2006)


\bibitem{BESIII:intro} M. Ablikim \etal, Nucl. Instrum. Methods Phys. Res. A, 614: 345 (2010)

\bibitem{bes3:KKMC} R.~G.~Ping , Chin. Phys. C, 32: 599 (2008)

\bibitem{Deng:2007zzb} Z.~Deng, H.~Liu, G.~Cao, M.~He, Y.~Yuan, Z.~You and Y.~Liang,
PoS ACAT, 043 (2007)


\bibitem{bes3:p1} M. Ablikim \etal~(BES Collaboration), Phys. Lett. B, 660: 315 (2008)

\bibitem{bes3:p2} M. Ablikim \etal~(BESIII Collaboration), Phys. Rev. D, 96: 032004 (2017)

\bibitem{bes3:p3} M. Ablikim \etal~(BESIII Collaboration), Phys. Rev. Lett., 118: 092001 (2017)

\bibitem{bes3:p4} M. Ablikim \etal~(BESIII Collaboration), Phys. Rev. Lett., 118: 092002 (2017).

\bibitem{bes3:p5} M. Ablikim \etal~(BESIII Collaboration), Phys. Rev. Lett., 122: 102002 (2019)



\bibitem{bes3:p6} M. Ablikim \etal~(BESIII Collaboration), Phys. Rev. Lett., 122: 232002 (2019)

\bibitem{bes3:p7} M. Ablikim \etal~(BESIII Collaboration), Phys. Rev. D, 99: 091103 (2019)


\bibitem{Actis:2010gg}
S.~Actis \textit{et al.}
Eur. Phys. J. C 66: 585 (2010)



\end{references}
\end{document}